\documentclass[a4paper,11pt]{article}
\usepackage{pos}

\title{Search for new phenomena in top quark interactions}
\author[a]{K. Skovpen (on behalf of the ATLAS and CMS Collaborations)}

\affiliation[a]{Ghent University,\\
  Proeftuinstraat 86, B-9000 Gent, Belgium}

\abstract{The study of the processes with the production of top quarks
represents a unique possibility to test the standard model (SM) predictions
and probe the new physics effects. Potential deviations from the SM
expectations are parametrized within the framework of the Effective Field
Theory (EFT). The latest EFT results from ATLAS and CMS experiments are presented. 
Dedicated studies of processes with flavour-changing neutral
currents are also discussed.
}

\FullConference{%
  The Eighth Annual Conference on Large Hadron Collider Physics-LHCP2020 \\
  25-30 May, 2020\\
  online}

\begin{document}
\maketitle

The top quark sector plays an important role in precision tests of the
standard model (SM) predictions and searches for new physics. The heaviness of the top quark and its large coupling
strength of the interaction with the Higgs boson suggests
an enhanced sensitivity to hypothetical new heavy particles and
interactions. The potential new physics effects can be probed in the
study of the SM processes with top quarks, as well as by searching for
forbidden interactions of these elementary particles. This report
summarizes the latest experimental results on these studies from the
ATLAS~\cite{PUB_ATLAS} and CMS~\cite{PUB_CMS} experiments using
proton-proton collision data collected at $\sqrt{s} = $~13~TeV at the LHC.

The Effective Field Theory (EFT) approach represents a generalized parametrization
of various new physics effects on the basis of dimension-six operators~\cite{PUB_EFT1, PUB_EFT2}.
The SMEFT framework provides a description of all possible interaction
vertices involving the SM elementary particles~\cite{PUB_SMEFT}. 
The processes involving the so-called flavour-changing neutral
currents (FCNC) are among possible anomalous interactions of top
quarks, which are also included in the EFT description~\cite{PUB_FCNC}. These
interactions are forbidden in the SM at tree level and are
significantly suppressed in higher orders~\cite{PUB_GIM}.

\begin{figure}[hbtp]
\begin{center}
\includegraphics[width=0.55\linewidth]{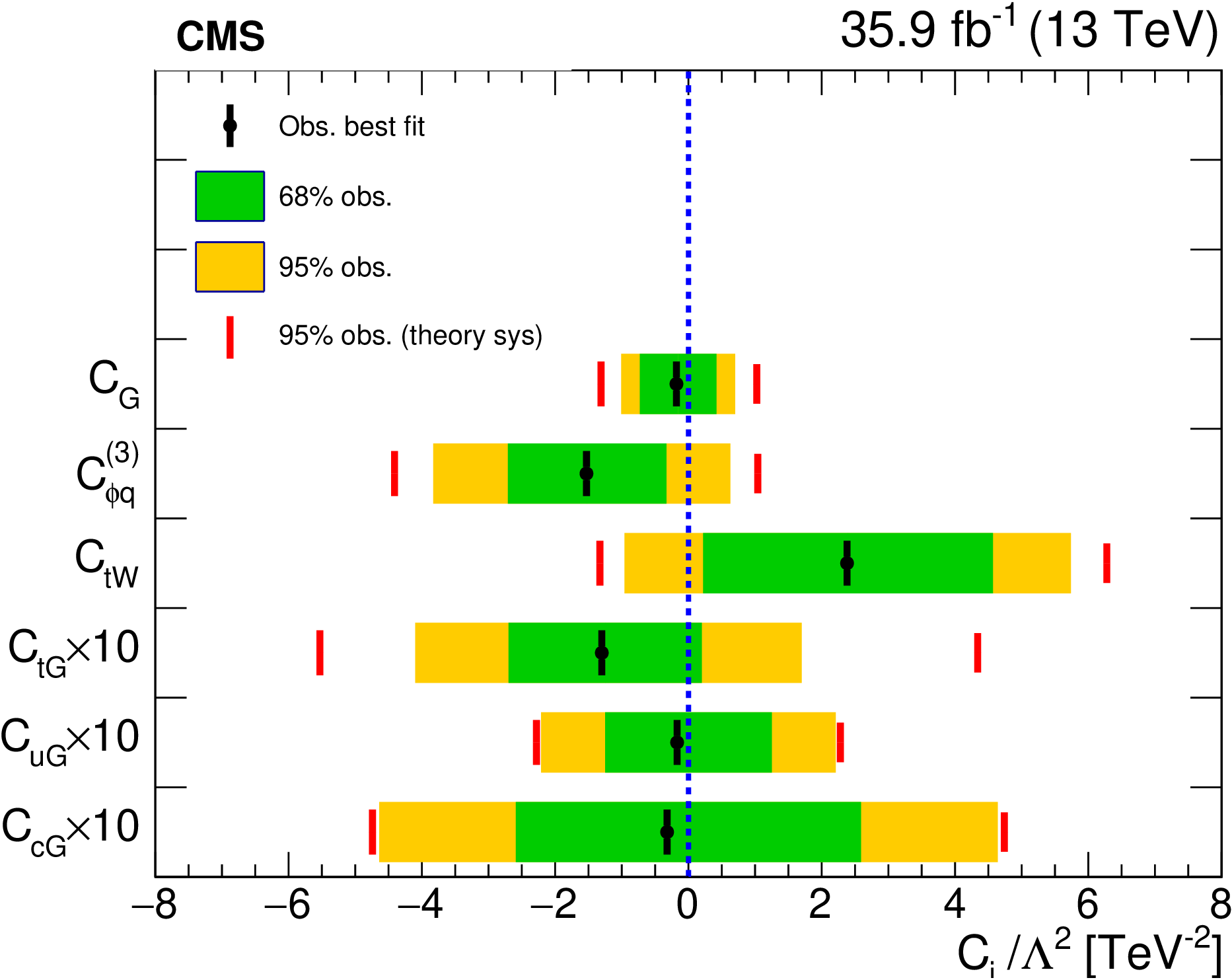}
\caption{The observed constraints on the EFT Wilson coefficients in the
analysis~\cite{PUB_2L}. The dashed line shows the SM expectation value.
The theory uncertainty here encodes the uncertainties on the
theoretical predictions for the signal cross section arising from the
choice of factorization and renormalization scales as well as parton
distribution functions. One non-vanishing Wilson coefficient is assumed at a time.
}
\label{fig:CMSEFT}
\end{center}
\end{figure}

Processes with the production of top quark pairs ($\mathrm{t\bar{t}}$), as well as the
associated production of a top quark with a W boson ($\mathrm{tW}$), are sensitive to
several EFT operators involving the top quark. The first simultaneous study of
these two production modes is performed in dilepton opposite-sign
events using 36 fb$^{-1}$ of the data collected by CMS~\cite{PUB_2L}. The analysis
probes the anomalous Wtb and FCNC couplings in the $\mathrm{tW}$ production mode,
while the triple gluon field interactions are studied in the
$\mathrm{t\bar{t}}$ production channel. Additionally, the chromomagnetic dipole
moments of the top quark are probed in both production modes. The
study of EFT effects is optimized using neural network (NN)
discriminants. The NN distributions are used to extract the constraints on the relevant EFT Wilson
coefficients through a likelihood fit to data of the expected signal
and background contributions (Fig.~\ref{fig:CMSEFT}).

The chromomagnetic and chromoelectric dipole moments of the top quark are studied in
the opposite-sign dilepton analysis of the $\mathrm{t\bar{t}}$ spin
correlations and differential production cross
sections with 36 fb$^{-1}$ of CMS data~\cite{PUB_CDM}. The full spin
density matrix is measured to probe various anomalous effects of the
four-particle effective couplings with top quarks. The resulting
constraints on the chromomagnetic and
chromoelectric moments are -0.24 <
$\mathrm{C_{tG}/\Lambda^{2}}$ < 0.07 TeV$^{-2}$ and -0.33 <
$\mathrm{C^{I}_{tG}/\Lambda^{2}}$ < 0.20 TeV$^{-2}$, respectively. The obtained results
represent a significant improvement over the previously published
direct limits.

A model independent search for top quark charged lepton flavour
violation (cLFV) decays is performed in the final states with three
leptons using 80 fb$^{-1}$ of the ATLAS data~\cite{PUB_LFV}. The analysis
is sensitive to the axial-vector, scalar, pseudo-scalar, and
lepton-quark EFT operators. A Boosted Decision Tree (BDT)
discriminant is used to suppress the SM backgrounds. The observed
(expected) resulting
constraints on the top quark cLFV branching fraction, $\mathrm{B(t \rightarrow
ll'q) < 1.9\ (1.4) \cdot 10^{-5}}$, improve over the best indirect limits by
several orders of magnitude.

The recent studies of the top quark FCNC interactions with a photon
($\gamma$) using
81 fb$^{-1}$ of the ATLAS data have led to significantly improved constraints on
these types of interaction~\cite{PUB_FCNCPHO}. The
analysis is performed in the single lepton final states with a photon. 
The FCNC effects are simultaneously probed in the top quark
decays in $\mathrm{t\bar{t}}$ events, as well as in the associated
production of the top quark with a photon via anomalous FCNC couplings. The NN discriminant is used
to suppress the $\mathrm{W/Z+\gamma}$ processes, as well as events
with mis-identified photons. The observed (expected) limits on the top quark FCNC
decay branching fraction for left- and right-handed couplings with an
up-quark are
$\mathrm{B(t \rightarrow u\gamma) < 2.8\ (4.0) \cdot 10^{-5}}$ and
$\mathrm{B(t \rightarrow u\gamma) < 6.1\ (5.9) \cdot 10^{-5}}$,
respectively. In the case of the top quark FCNC couplings with a charm
quark, the corresponding limits are
$\mathrm{B(t \rightarrow c\gamma) < 22\ (27) \cdot 10^{-5}}$
$\mathrm{B(t \rightarrow c\gamma) < 16\ (28) \cdot 10^{-5}}$,
respectively.

\begin{figure}[hbtp]
\begin{center}
\includegraphics[width=0.60\linewidth]{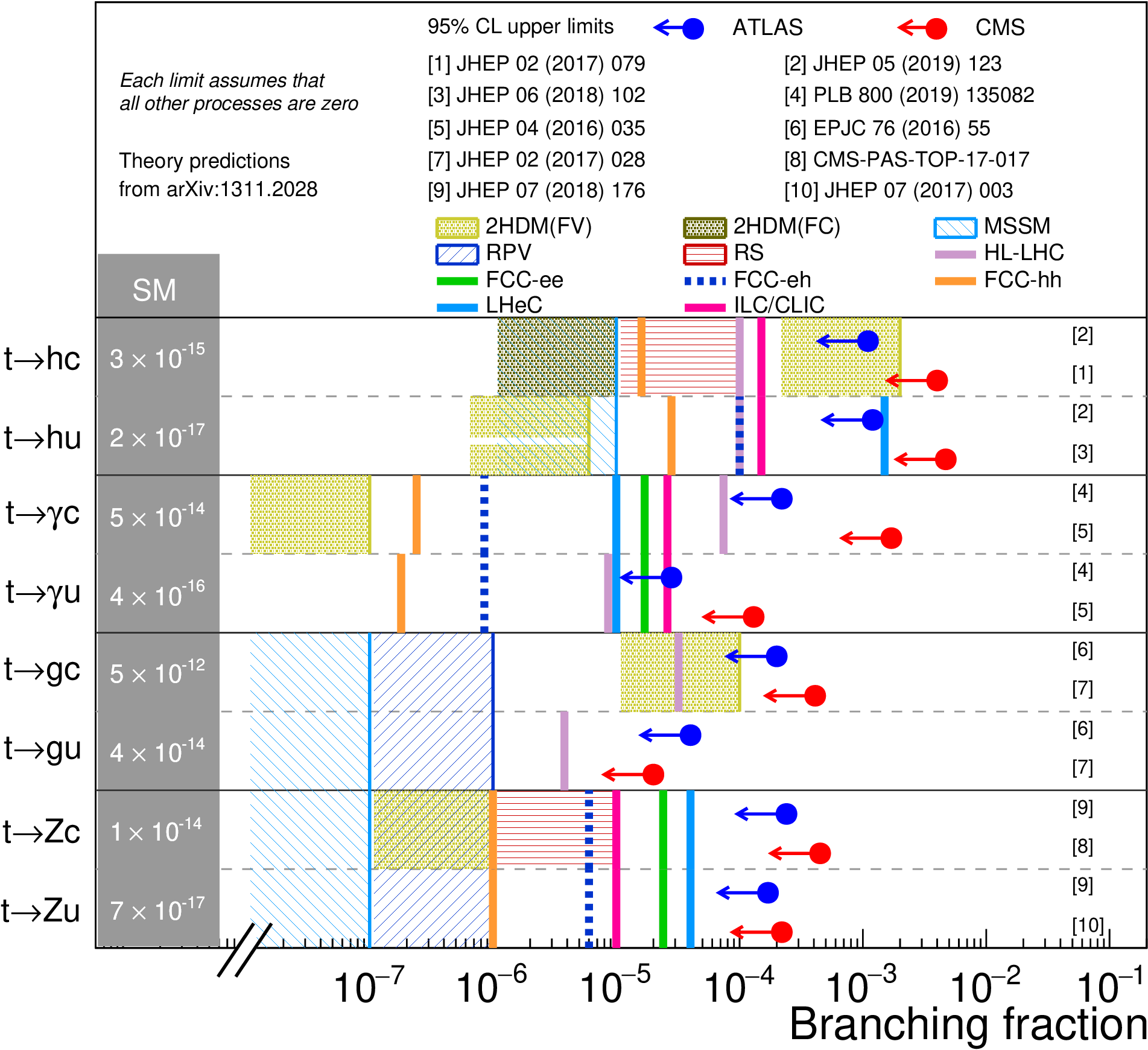}
\caption{
The updated summary of the FCNC searches with top quarks from
Ref.~\cite{PUB_FCNC2}.
}
\label{fig:FCNC}
\end{center}
\end{figure}

The combination of the ATLAS results obtained from the studies of various final
states relevant to the top quark FCNC interactions with the Higgs
boson ($\mathrm{h}$) is
performed~\cite{PUB_FCNCHIG} with 36 fb$^{-1}$ of recorded data. The considered final states arise from
the Higgs boson decays to two b quarks, two tau leptons, WW and ZZ
boson pairs, and two photons. The obtained observed (expected) limits
on the top FCNC decays via a Higgs boson with an up and a charm
quark are $\mathrm{B(t \rightarrow uh) < 1.2\ (0.8) \cdot 10^{-3}}$ and
$\mathrm{B(t \rightarrow ch) < 1.1\ (0.8) \cdot 10^{-3}}$,
respectively. These results represent the best limits to date
on top-Higgs FCNC interactions.

The summary of the latest constraints obtained with the LHC data on the
top quark FCNC branching ratios, along
with the theoretical predictions~\cite{PUB_FCNC} and projections of sensitivity of
experimental searches for future experiments~\cite{PUB_FCNC2,
PUB_FUTSM, PUB_FUTFCNC}, is presented in
Fig.~\ref{fig:FCNC}. The shown experimental limits represent an
important probe of various models of new physics.

The presented experimental results on the EFT and FCNC searches with
top quarks show a great improvement over the previously published
results, as well as introduce novel approaches of studying the top
quark anomalous interactions. The described results were obtained using the partial data collected
during the second run of the data taking at the LHC, and further
updates from the analysis of the full data are foreseen in the future.


\begin{thebibliography}{99}
\bibitem{PUB_ATLAS} ATLAS Collaboration, JINST 3 (2008) S08003.
\bibitem{PUB_CMS} CMS Collaboration, JINST 3 (2008) S08004.
\bibitem{PUB_EFT1} W. Buchmuller and D. Wyler, Nucl. Phys. B 268 (1986) 621.
\bibitem{PUB_EFT2} B. Grzadkowski, M. Iskrzynski, M. Misiak et al., JHEP 10 (2010) 085.
\bibitem{PUB_SMEFT} J. A. Aguilar-Saavedra et al., arXiv:1802.07237.
\bibitem{PUB_FCNC} J. A. Aguilar-Saavedra, Acta Phys. Polon. B 35 (2004) 2695.
\bibitem{PUB_GIM} S. L. Glashow, J. Iliopoulos, and L. Maiani, Phys. Rev. D 2 (1970) 1285.
\bibitem{PUB_2L} CMS Collaboration, Eur. Phys. J. C 79 (2019) 886.
\bibitem{PUB_CDM} CMS Collaboration, Phys. Rev. D 100 (2019) 072002.
\bibitem{PUB_LFV} ATLAS Collaboration, ATLAS-CONF-2018-044, http://cds.cern.ch/record/2638305.
\bibitem{PUB_FCNCPHO} ATLAS Collaboration, Phys. Lett. B 800 (2019) 135082.
\bibitem{PUB_FCNCHIG} ATLAS Collaboration, JHEP 05 (2019) 123.
\bibitem{PUB_FCNC2} A. Abada et al., Eur. Phys. J. C 79 (2019) 474.
%\bibitem{PUB_FCNC3} CMS Collaboration, JHEP 06 (2018) 102.
%\bibitem{PUB_FCNC4} CMS Collaboration, CMS PAS TOP-17-017, http://cds.cern.ch/record/2292045.
\bibitem{PUB_FUTSM} P. Azzi et al., arXiv:1902.04070.
\bibitem{PUB_FUTFCNC} A. Cerri et al., arXiv:1812.07638.

\end{thebibliography}
\end{document}